\documentclass[12pt]{article}
\usepackage[a4paper, left=2.5cm, right=2.5cm, top=2.5cm, bottom=2.5cm]{geometry} 
\usepackage{amsmath,amssymb,graphicx}
\usepackage{url}
\usepackage[T1]{fontenc}
\usepackage{lmodern}
\usepackage[utf8]{inputenc}
\usepackage{subcaption}
\usepackage{makecell}
\usepackage{booktabs}
\usepackage{placeins}

\begin{document}

\title{A Fibonacci-Based Ontogenetic Discretization of Body-Mass Growth Trajectories} 
\author{
Dorilson Silva Cambui\\
Governo de Mato Grosso, Secretaria de Estado de Educação\\
Cuiabá, Mato Grosso, Brazil\\
\texttt{dorilson.cambui@edu.mt.gov.br}\\
\texttt{dcambui@fisica.ufmt.br}
}

\maketitle

\begin{abstract}
\noindent	
	Body-mass growth is usually described by continuous models that represent the gradual increase of mass toward an adult or asymptotic value. In this work, we formulate and test an alternative description based on a Fibonacci-inspired ontogenetic coordinate. The model represents growth through a continuous or discretized developmental index, allowing the trajectory to be organized into fine substages. The main analysis compares the Fibonacci-based model with the ontogenetic growth model of West, Brown, and Enquist using digitized growth data from four species with markedly different body sizes and developmental times. The Fibonacci-based formulation was favored in two of the four cases, while the WBE equation remained favored in the other two. A complementary analysis using rodent growth data showed that the Fibonacci discretized model can also approach the performance of classical continuous models in some datasets, although not uniformly. These results suggest that a Fibonacci-based ontogenetic coordinate can provide a useful alternative representation of body-mass growth within a biological systems framework.
\end{abstract}

\section{Introduction}

The growth of living organisms is usually described as a continuous process. As time passes, body mass increases gradually and tends to approach a characteristic adult value. This behavior is commonly represented by classical growth functions, such as the Gompertz, logistic, von Bertalanffy, and Richards models \cite{gompertz1825,bertalanffy1957,richards1959,tjorve2010}. These models are useful because they describe a simple and intuitive idea. The organism gains mass rapidly during the early stages of development, but this gain becomes slower as the system approaches a limiting size \cite{bertalanffy1957,west2001}. From a physical point of view, this type of growth resembles a relaxation process. A system initially far from its final state changes rapidly, while a system closer to its final state changes more slowly. In biological growth, the final state is not a thermodynamic equilibrium in the strict sense. Even so, the analogy is useful because the mass trajectory reflects a balance between energy intake, internal organization, maintenance, and the production of new biomass. Although body mass is generally treated as a continuous variable, biological development is often discussed in terms of stages \cite{gilbert2000,minelli2003}. 

Among the theoretical models of ontogenetic growth, the model proposed by West, Brown, and Enquist occupies an important place \cite{west2001}. This model is based on the allocation of metabolic energy between maintenance of existing tissue and the production of new tissue. It provides a biologically motivated equation for growth and has been applied to organisms with very different body masses and developmental times. In their original work, West, Brown, and Enquist showed examples of growth curves for guinea pig, guppy, hen, and cow, illustrating the broad range of systems that can be described by the same theoretical framework. 

In a previous study, a different but related perspective was proposed using Fibonacci dynamics to describe metabolic scaling during growth \cite{cambui2026}. In that work, body mass and growth stage were connected through a Fibonacci-based scaling rule. The relation between mass and stage was used to define a dynamic metabolic exponent, written as a function of an ontogenetic index \(n\). Thus, instead of treating the metabolic exponent as a fixed quantity, the previous model allowed it to vary along the developmental trajectory. The central idea of that previous formulation was that the stage of growth can be represented by a mathematical index. This index does not need to be interpreted as a rigid biological step. Rather, it provides a way to organize development along a scale. In the Fibonacci-based model, the golden ratio enters as the scaling factor that connects consecutive stages of growth. This makes it possible to describe the organism not only by its chronological time, but also by its position along an ontogenetic trajectory. 

The present work uses this Fibonacci-based stage variable from a different point of view. The focus is no longer the metabolic exponent itself, but the growth curve of body mass. We ask whether a Fibonacci-inspired ontogenetic coordinate can reproduce growth trajectories that are usually described by continuous models. In particular, we test whether a discretized version of this coordinate can approximate empirical growth curves without assuming that body mass truly changes through abrupt biological jumps. This distinction is important. The proposed model does not claim that mass accumulation occurs by discontinuous jumps. Body mass may remain a continuous biological variable. The discretization is applied to the ontogenetic coordinate used to represent development. 
%In this sense, the model can be viewed as a stage-like or coarse-grained description of a continuous growth process.

The main objective of this study is to compare the Fibonacci-based model with the ontogenetic growth model of West, Brown, and Enquist. For this purpose, we use the four examples shown in their original work, corresponding to guinea pig, guppy, hen, and cow. These species are useful for the present test because they differ strongly in body size, growth time, and biological organization. Therefore, they provide a direct way to evaluate whether the Fibonacci-based description can follow growth trajectories across very different organisms. In addition to this main comparison, we also perform a shorter complementary test using rodent growth data. The rodent analysis includes body-mass trajectories for C57BL/6J mice, Wistar rats, and Sprague Dawley rats. This additional test is not intended to replace the comparison with the WBE model. Its purpose is to examine whether the same Fibonacci-based description remains useful when applied to independent growth series and when compared with classical continuous models. 

The results obtained in this work show that the Fibonacci-based model can provide a competitive description of ontogenetic growth in several cases. In the direct comparison with the WBE model, the Fibonacci-based formulation was favored for Guppy and Hen, while the WBE equation remained favored for Guinea pig and Cow. 
Even in the cases where WBE was favored, the Fibonacci curves followed the empirical trajectories closely.
The complementary rodent analysis also showed that the Fibonacci discretized model can approach the performance of classical continuous models in some growth series, although not in all of them. Thus, the main result of this study is not that Fibonacci dynamics replaces established growth equations. Rather, it shows that a Fibonacci-inspired ontogenetic coordinate can represent body-mass growth with reasonable accuracy across different biological systems.

\section{Model formulation}

\noindent\textbf{The WBE ontogenetic growth model.}
The first model considered in this work is the ontogenetic growth model proposed by West, Brown, and Enquist. 
This model is based on a physical and biological idea. 
During growth, metabolic energy is divided between two main processes. 
One part is used to maintain the existing tissues, while another part is used to produce new biomass.
In this framework, body mass growth is described by the equation

\begin{equation}
	\frac{dm}{dt}
	=
	a m^{3/4}
	\left[
	1-
	\left(
	\frac{m}{M}
	\right)^{1/4}
	\right],
\end{equation}
where \(m(t)\) is the body mass at time \(t\), \(M\) is the asymptotic body mass, and \(a\) is a growth parameter related to the rate at which metabolic energy is transformed into new tissue. The integrated form of this model is

\begin{equation}
	\left(
	\frac{m}{M}
	\right)^{1/4}
	=
	1-
	\left[
	1-
	\left(
	\frac{m_0}{M}
	\right)^{1/4}
	\right]
	\exp
	\left(
	-\frac{a t}{4M^{1/4}}
	\right),
\end{equation}
where \(m_0\) is the initial body mass. Equivalently, body mass can be written as

\begin{equation}\label{west} 
	m(t)
	=
	M
	\left\{
	1-
	\left[
	1-
	\left(
	\frac{m_0}{M}
	\right)^{1/4}
	\right]
	\exp
	\left(
	-\frac{a t}{4M^{1/4}}
	\right)
	\right\}^{4}.
\end{equation}

This equation gives a continuous growth curve. 
At early times, the organism grows rapidly. 
As \(m(t)\) approaches \(M\), the growth rate decreases. 
This behavior is consistent with the general idea that growth slows down as more metabolic energy is required for maintenance.

\noindent\textbf{Fibonacci-based ontogenetic coordinate.}
The second model considered in this work is based on a Fibonacci-inspired ontogenetic coordinate. 
This approach was motivated by a previous study in which body mass and growth stage were related through a Fibonacci-based scaling rule~\cite{cambui2026}.  In that study, the stage variable was used to define a dynamic metabolic exponent during growth.
The starting point is the relation

\begin{equation}
	M = M_0\phi^n,
\end{equation}
where \(M_0\) is a reference mass, \(n\) is an ontogenetic index, and

\begin{equation}
	\phi=
	\frac{1+\sqrt{5}}{2}
\end{equation}
is the golden ratio.
The inverse relation is

\begin{equation}
	n=
	\log_{\phi}
	\left(
	\frac{M}{M_0}
	\right).
\end{equation}
This expression allows the stage index \(n\) to be obtained from the relative body mass \(M/M_0\). 
In the present work, this same idea is used to describe the body-mass trajectory itself. 
The variable \(n\) is interpreted as an ontogenetic coordinate that measures the progression of the organism along its growth trajectory.

The use of the golden ratio is not introduced here as an arbitrary numerical fitting constant. It comes from the previous Fibonacci-based metabolic scaling framework, in which consecutive ontogenetic stages were associated with a recursive growth structure. In the asymptotic limit of the Fibonacci sequence, the ratio between consecutive terms tends to the golden ratio. Thus, \(\phi\) appears as the natural scaling factor of a recursive stage-based representation of growth. In the present work, this idea is not used to claim that biological growth is exactly Fibonacci. Rather, \(\phi\) defines a fixed scale factor that connects the ontogenetic coordinate with body mass, allowing the model to be tested against empirical growth trajectories. Other scale factors could be mathematically tested, but the purpose of the present study is to evaluate the specific ontogenetic coordinate derived from the Fibonacci-based metabolic framework.

To describe growth as a function of time, the ontogenetic coordinate is written as \(n(t)\). 
The continuous Fibonacci-based model is then given by

\begin{equation}\label{Mn}
	m(t)=m_0\phi^{n(t)},
\end{equation}
where \(m_0\) is the first observed mass of the organism or group being analyzed.
The function \(n(t)\) is assumed to have a saturating form,

\begin{equation}\label{n-infinito}
	n(t)
	=
	n_{\infty}
	\left[
	1-
	\exp
	\left(
	-k(t-t_0)
	\right)
	\right],
\end{equation}
where \(n_{\infty}\) is the limiting value of the ontogenetic coordinate, \(k\) controls the rate of progression along the trajectory, and \(t_0\) is a temporal shift.
This choice preserves the physical idea of growth with saturation. 
At the beginning of development, the ontogenetic coordinate changes more rapidly. 
At later times, its variation becomes slower as the organism approaches a mature state.

\noindent\textbf{Relation between the continuous Fibonacci model and Gompertz growth.}
The continuous Fibonacci-based model can be related to the Gompertz growth equation. 
Starting from

\begin{equation}
	m(t)=m_0\phi^{n(t)}
\end{equation}
and substituting the saturating expression for \(n(t)\), we obtain

\begin{equation}
	m(t)
	=
	m_0
	\phi^{
		n_{\infty}
		\left[
		1-
		\exp
		\left(
		-k(t-t_0)
		\right)
		\right]
	}.
\end{equation}
Using the identity \(\phi^x=\exp(x\ln\phi)\), this can be written as

\begin{equation}
	m(t)
	=
	m_0
	\exp
	\left\{
	n_{\infty}\ln\phi
	\left[
	1-
	\exp
	\left(
	-k(t-t_0)
	\right)
	\right]
	\right\}.
\end{equation}
After rearrangement, this expression becomes

\begin{equation}
	m(t)
	=
	m_0\phi^{n_{\infty}}
	\exp
	\left[
	-
	n_{\infty}\ln\phi
	\,
	e^{kt_0}
	e^{-kt}
	\right].
\end{equation}
Defining

\begin{equation}
	A=m_0\phi^{n_{\infty}}
\end{equation}
and

\begin{equation}
	B=n_{\infty}\ln\phi \, e^{kt_0},
\end{equation}
we obtain

\begin{equation}\label{gomp}
	m(t)=A\exp\left[-B\exp(-kt)\right],
\end{equation}
which has the classical mathematical form of the Gompertz growth model.
Here, \(A\) is the asymptotic mass, \(B\) is related to the initial condition, and \(k\) controls the growth rate.

Therefore, the continuous Fibonacci-based formulation should be interpreted as a Gompertz-type reparametrization written in terms of an ontogenetic coordinate.  The original element of the present approach is not the continuous curve alone, but the use and discretization of the ontogenetic coordinate \(n(t)\).

\noindent\textbf{Discretized Fibonacci-based growth model.}
To test a stage-like representation of growth, the continuous ontogenetic coordinate \(n(t)\) is discretized into substages. 
The discretized coordinate is defined as

\begin{equation}\label{deltan}
	n_{\Delta}(t)
	=
	\Delta n
	\,
	\mathrm{round}
	\left[
	\frac{n(t)}{\Delta n}
	\right],
\end{equation}
where \(\Delta n\) is the size of the ontogenetic substage. The corresponding discretized Fibonacci model is

\begin{equation}\label{mfibdisc}
	m(t)=m_0\phi^{n_{\Delta}(t)}.
\end{equation}
Combining these expressions, the discretized Fibonacci-based model can be written as
\begin{equation}
	m_{\Delta}(t)
	=
	m_0
	\phi^{
		\Delta n
		\,
		\mathrm{round}
		\left[
		\frac{
			n_{\infty}
			\left[
			1-\exp\left(-k(t-t_0)\right)
			\right]
		}
		{\Delta n}
		\right]
	}.
	\label{fibdiscfull}
\end{equation}
Equation~\ref{fibdiscfull} is the explicit form of the discretized Fibonacci model used in the numerical fits.

In this formulation, the discretization is applied to the ontogenetic coordinate, not directly to body mass. 
Thus, the step-like structure of the model should not be interpreted as evidence that body mass grows through abrupt biological jumps. 
There is no biological evidence for such a claim in the present analysis. Instead, the step-like structure represents a coarse-grained description of developmental progression. Small values of \(\Delta n\) produce a fine discretization and make the curve closer to the continuous model. Larger values of \(\Delta n\) produce a more visible step-like structure. In this sense, \(\Delta n\) controls the resolution of the ontogenetic scale.
The parameter \(\Delta n\) is treated here as an effective resolution parameter. It is not assumed to be a universal biological constant. 
Different values of \(\Delta n\) may reflect differences in growth pattern, sampling interval, data resolution, or the degree to which a discretized ontogenetic coordinate can approximate the observed trajectory. 
%For this reason, the discretized Fibonacci model is interpreted as an alternative representation of growth rather than as a replacement for classical continuous models. Its relevance is evaluated by comparing its performance with established growth models, especially the WBE ontogenetic growth model.

\section{Methods} %%%%%%%%%%%%%%%%%%%%%%%%%%%%%%%%%%%%%%%%%%%%%%%%%%%%%%%%%%%%%%%%%%%%%%%%%%%%%%%%%%%%%%%%%%%%%%%%%%%%%

\noindent\textbf{Data used in the WBE comparison.}
The main analysis was based on the four growth examples presented by West, Brown, and Enquist in their ontogenetic growth study. 
The organisms considered were guinea pig, guppy, hen, and cow. 
These species were chosen because they represent very different biological systems, with distinct body masses, developmental times, and growth patterns. The empirical points were obtained by digitizing the growth curves shown in Fig.~1 of West, Brown, and Enquist \cite{west2001}. 
The digitization was performed separately for each panel, using the axes shown in the original figure. 
The resulting data consisted of pairs $(t_i,m_i)$, where \(t_i\) is time and \(m_i\) is the corresponding body mass.
Because these data were extracted from a published figure, the comparison should be interpreted as an exploratory analysis. 
The purpose was not to replace the original data sources used by West, Brown, and Enquist, but to test whether the Fibonacci-based model can reproduce the same type of growth trajectories used in their illustrative examples.

As mentioned in the Introduction, the WBE model was used as the main reference model in this study. 
This model describes growth from the allocation of metabolic energy between the maintenance of existing tissue and the production of new biomass. The integrated form used in the fitting procedure is given by Equation~\ref{west}. 
In the numerical fitting, the WBE model was treated as a three-parameter model, with \(a\), \(M\), and \(m_0\) estimated from the digitized data. This allowed a direct comparison with the Fibonacci-based models using the same empirical points.

\noindent\textbf{Fibonacci-based continuous model.}
The continuous Fibonacci-based model was constructed from an ontogenetic coordinate \(n(t)\). 
Body mass was written as \(m(t)=m_0\phi^{n(t)}\), as given in Equation~\ref{Mn}.
The ontogenetic coordinate was modeled as a saturating function   $n(t)=n_{\infty}\left[1-\exp\left(-k(t-t_0)\right)\right]$, as given in Equation~\ref{n-infinito}, where \(n_{\infty}\) is the limiting ontogenetic coordinate, \(k\) controls the rate of progression along the growth trajectory, and \(t_0\) is a temporal shift.

In this formulation, \(m_0\) was fixed as the first observed body mass of each dataset. 
Therefore, it was not counted as a fitted parameter. 
The continuous Fibonacci-based model was treated as a three-parameter model, with fitted parameters \(n_{\infty}\), \(k\), and \(t_0\).

\noindent\textbf{Fibonacci-based discretized model.} 
To test a stage-like representation of growth, the continuous ontogenetic coordinate was discretized into substages. 
The discretized coordinate was defined by Equation~\ref{deltan}, where \(\Delta n\) is the size of the ontogenetic substage. 
The discretized Fibonacci-based growth model was then written as Equation~\ref{mfibdisc}.

In this formulation, the discretization is applied to the ontogenetic coordinate, not directly to body mass. 
Thus, the step-like structure of the model should not be interpreted as evidence that body mass grows through abrupt biological jumps. 
There is no biological evidence for such a claim in the present analysis. 
Instead, the step-like structure represents a coarse-grained description of developmental progression.

Small values of \(\Delta n\) produce a fine discretization and make the curve closer to the continuous model. 
Larger values of \(\Delta n\) produce a more visible step-like structure. 
In this sense, \(\Delta n\) controls the resolution of the ontogenetic scale.

For the WBE comparison, \(\Delta n\) was selected by a grid search over the interval
\[
0.01 \leq \Delta n \leq 0.50.
\]
The value that minimized the RMSE of the discretized Fibonacci model was selected for each dataset. 
The same criterion was also applied to the rodent datasets. 
Because \(\Delta n\) was selected from the data, the discretized Fibonacci model was treated as having four effective parameters, namely \(n_{\infty}\), \(k\), \(t_0\), and \(\Delta n\).

In the final analysis, the best established continuous model was selected using \(AIC_c\). 
Thus, the model-selection criteria were applied consistently across the datasets.

A possible limitation of this procedure is that \(\Delta n\) is selected using the same data employed for model fitting. 
Thus, the grid search may favor the discretized Fibonacci model by choosing the value of \(\Delta n\) that best adapts to the observed points. 
Although \(\Delta n\) was treated as an additional effective parameter in the information criteria, this does not completely remove the risk of overfitting. 
For this reason, the values of \(\Delta n\) should be interpreted as effective resolution parameters of the fitted trajectories, not as intrinsic biological constants.

\noindent\textbf{Additional rodent datasets.} %%%%%%%%%%%%%%%%%%%%%%%%%%%%%%%%%%%%%%%%%%%%%%%%%%%%%%%%%%%%%%%%%%%%%%%%%%%%%%%%%%%%%%
As a complementary analysis, the Fibonacci-based model was also tested using rodent growth data. 
The datasets included six growth series: C57BL/6J males, C57BL/6J females, Wistar males, Wistar females, Sprague Dawley males, and Sprague Dawley females. 
The C57BL/6J data, covering males and females from 3 to 24 weeks of age, were obtained from body-weight information provided by The Jackson Laboratory \cite{jax_c57bl6j_body_weight}. 
The Wistar and Sprague Dawley data, also separated by sex and covering the interval from 3 to 12 weeks of age, were obtained from an age-in-week body-weight table provided by the Laboratory Animal Center of National Cheng Kung University \cite{ncku_rat_body_weight}.

The rodent data were used to evaluate whether the Fibonacci-based discretized model also performs reasonably when applied to independent growth series. 
In this complementary analysis, the Fibonacci-based model was compared with classical continuous growth models, including the Gompertz equation, the logistic equation, the von Bertalanffy equation, and the Richards equation. 
This rodent analysis was not treated as the main validation of the model. 
Rather, it was used as an additional test of consistency, showing how the proposed ontogenetic coordinate behaves in a different set of growth data.

\noindent\textbf{Classical continuous growth models for the rodent analysis.}
For the rodent datasets, four classical continuous models were used as references. 
The Gompertz model was defined as in Equation~\ref{gomp}.
The logistic model was written as

\begin{equation}
	m(t)=\frac{A}{1+B\exp(-kt)}.
\end{equation}
The von Bertalanffy model was written as

\begin{equation}
	m(t)=A\left[1-B\exp(-kt)\right]^3.
\end{equation}
The Richards model was written as

\begin{equation}
	m(t)=
	\frac{A}
	{\left[1+B\exp(-kt)\right]^{1/\nu}},
\end{equation}
where \(\nu\) is an additional shape parameter. 
This parameter gives the Richards model greater flexibility than the logistic model.

All models were fitted by nonlinear least squares. 
For each rodent dataset, the best established continuous model was selected using the corrected Akaike information criterion.

\noindent\textbf{Error measures and information criteria.}
Model performance was evaluated using standard error measures and information criteria.
The root mean square error was defined as

\begin{equation}
	RMSE=
	\sqrt{
		\frac{1}{N}
		\sum_{i=1}^{N}
		\left[
		m_i-m_{\mathrm{model}}(t_i)
		\right]^2
	},
\end{equation}
where \(m_i\) is the observed body mass, \(m_{\mathrm{model}}(t_i)\) is the value predicted by the model, and \(N\) is the number of observations. The mean absolute error was defined as

\begin{equation}
	MAE=
	\frac{1}{N}
	\sum_{i=1}^{N}
	\left|
	m_i-m_{\mathrm{model}}(t_i)
	\right|.
\end{equation}
The sum of squared errors was defined as

\begin{equation}
	SSE=
	\sum_{i=1}^{N}
	\left[
	m_i-m_{\mathrm{model}}(t_i)
	\right]^2.
\end{equation}
The Akaike information criterion (AIC) and the Bayesian information criterion (BIC) were used to compare the fitted models while considering the number of fitted parameters \cite{akaike1974,schwarz1978,burnham2002}. The Akaike information criterion  was computed as 

\begin{equation}
	AIC=
	N\ln\left(\frac{SSE}{N}\right)+2p,
\end{equation}
and the Bayesian information criterion  was computed as

\begin{equation}
	BIC=
	N\ln\left(\frac{SSE}{N}\right)+p\ln(N),
\end{equation}
where \(p\) is the number of fitted or effective model parameters.

Because some datasets contain a limited number of observations, especially in the complementary rodent analysis, the corrected Akaike information criterion was also computed. 
The corrected criterion was defined as

\begin{equation}
	AIC_c
	=
	AIC
	+
	\frac{2p(p+1)}{N-p-1},
	\label{aicc}
\end{equation}
where \(N\) is the number of observations and \(p\) is the number of fitted or effective parameters. 
The \(AIC_c\) correction is especially relevant when \(N\) is small relative to \(p\). 
When \(N\leq p+1\), \(AIC_c\) is not defined, and this limitation must be considered in very small fitting windows.

The AIC, \(AIC_c\), and BIC were used because they penalize models with more parameters. 
The corrected criterion \(AIC_c\) was included to provide a more conservative comparison for small datasets.
This is important in the present study because the discretized Fibonacci model includes the additional effective parameter \(\Delta n\).

\noindent\textbf{Numerical implementation.}
All numerical analyses were performed using Python. 
Nonlinear curve fitting was performed with the \texttt{curve\_fit} routine from the \texttt{scipy.optimize} library. 
Data handling was performed using \texttt{numpy} and \texttt{pandas}. 
Figures were generated using \texttt{matplotlib}.
For each dataset, the WBE model, the continuous Fibonacci model, and the discretized Fibonacci model were fitted to the same empirical points. 
The resulting curves were compared using RMSE, MAE, SSE, AIC, \(AIC_c\), and BIC.
For the rodent datasets, the same procedure was applied, with the addition of the classical continuous models described above. 
For the rodent datasets, \(\Delta n\) was also selected by grid search over the same interval, using the value that minimized the RMSE of the discretized Fibonacci model for each dataset.

\section{Results}

\noindent\textbf{Comparison with the WBE ontogenetic growth model.}
The first analysis compared the WBE ontogenetic growth model with the continuous and discretized Fibonacci-based models. 
The comparison was performed using digitized growth data for guinea pig, guppy, hen, and cow. 
These examples were chosen because they correspond to organisms with very different body masses and developmental times.

\begin{figure}[htbp]
	\centering
	\includegraphics[width=0.99\textwidth]{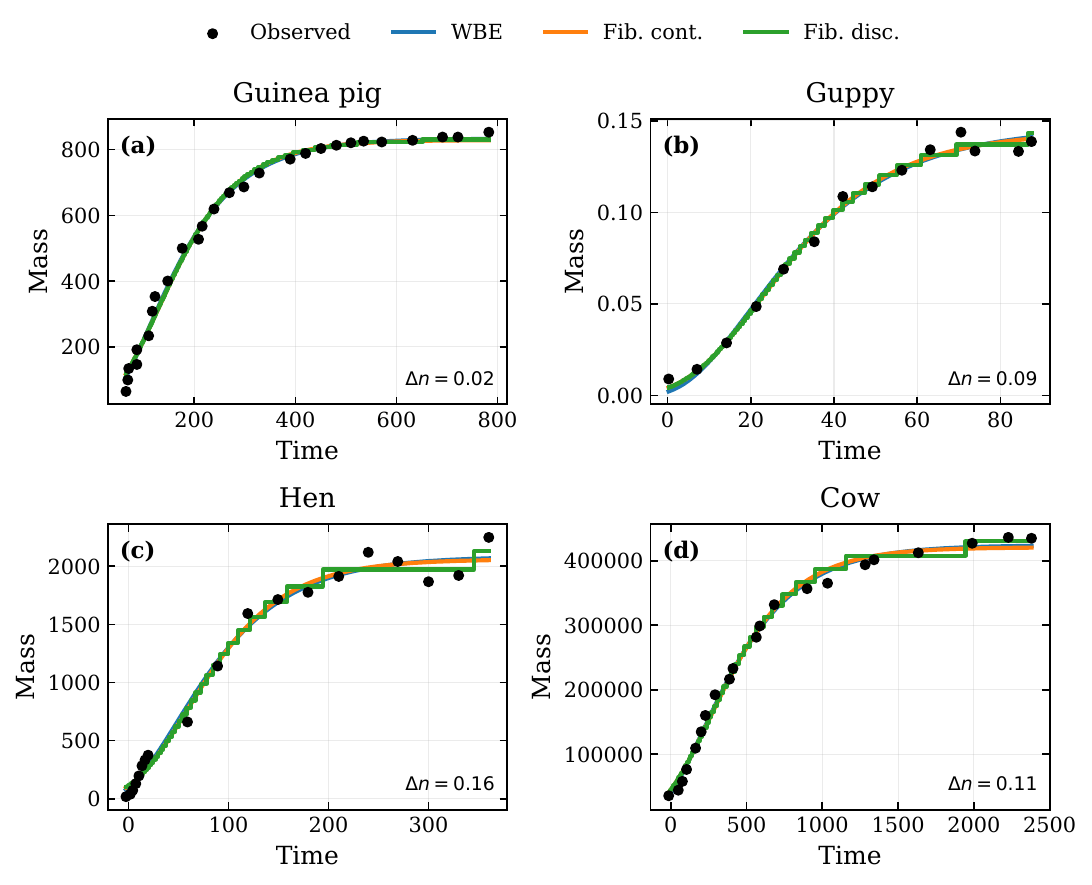}
	\caption{
		Comparison between the WBE ontogenetic growth model and the Fibonacci-based models for four species. 
		The black points represent digitized empirical data. 
		The blue curve represents the WBE equation, the orange curve represents the continuous Fibonacci model, and the green step-like curve represents the discretized Fibonacci model. 
		The discretized curve should be interpreted as a coarse-grained representation of the ontogenetic coordinate, not as evidence of abrupt jumps in body mass. Empirical points were digitized from Fig.~1 of West, Brown, and Enquist \cite{west2001}.
	}
	\label{fig:wbe_fibonacci}
\end{figure}

Figure~\ref{fig:wbe_fibonacci} shows the fitted curves for the four species. The empirical points were obtained by digitizing the growth curves shown in Fig.~1 of West, Brown, and Enquist \cite{west2001}. 
This comparison should be interpreted as an exploratory analysis based on digitized data.
The WBE equation, the continuous Fibonacci model, and the discretized Fibonacci model were all able to reproduce the general shape of the observed growth trajectories. 
In particular, the Fibonacci-based curves followed the rapid initial increase in body mass and the later tendency toward saturation.

The quantitative comparison is shown in Table~\ref{tab:wbe_fibonacci_aicc}. 
For guinea pig and cow, the WBE equation produced the best support according to the information criteria. 
For guppy, the discretized Fibonacci model produced the lowest RMSE, AIC, and BIC, whereas the continuous Fibonacci model produced a slightly lower \(AIC_c\). 
Thus, the corrected criterion favored the Fibonacci formulation, but not specifically its discretized version in this dataset. 
For hen, the discretized Fibonacci model produced the lowest RMSE, AIC, \(AIC_c\), and BIC values.

Overall, the Fibonacci-based formulation was favored in two of the four cases, while the WBE model was favored in the other two. 
This result indicates that the Fibonacci approach can be competitive with the WBE model, although its advantage depends on the dataset and on the information criterion considered. 
Therefore, the Fibonacci-based formulation should not be interpreted as a universal replacement for the WBE model. 
Rather, it provides an alternative ontogenetic parametrization that can be competitive in some biological systems.

\begin{table}[htbp]
	\centering
	\caption{
		Error measures and information criteria for the WBE and Fibonacci-based models. 
		Lower AIC, \(AIC_c\), and BIC values indicate better support after penalizing model complexity. 
		The discretized Fibonacci model was treated as a four-parameter model because \(\Delta n\) was selected from the data.
	}
	\label{tab:wbe_fibonacci_aicc}
	\begin{tabular}{l l c c c c c}
		\hline
		Species & Model & \(p\) & RMSE & AIC & \(AIC_c\) & BIC \\
		\hline
		Guinea pig & WBE & 3 & 20.1196 & 168.0916 & 169.1351 & 171.9791 \\
		Guinea pig & Fib. cont. & 3 & 22.9353 & 175.1646 & 176.2081 & 179.0522 \\
		Guinea pig & Fib. disc. & 4 & 22.8249 & 176.9041 & 178.7223 & 182.0875 \\
		\hline
		Guppy & WBE & 3 & 0.004505 & -145.2724 & -142.8724 & -143.3553 \\
		Guppy & Fib. cont. & 3 & 0.004006 & -148.5592 & -146.1592 & -146.6420 \\
		Guppy & Fib. disc. & 4 & 0.003492 & -150.4055 & -145.9610 & -147.8492 \\
		\hline
		Hen & WBE & 3 & 95.8855 & 179.3999 & 180.9999 & 182.2332 \\
		Hen & Fib. cont. & 3 & 95.7287 & 179.3377 & 180.9377 & 182.1710 \\
		Hen & Fib. disc. & 4 & 85.5057 & 177.0461 & 179.9033 & 180.8239 \\
		\hline
		Cow & WBE & 3 & 9894.0898 & 392.3871 & 393.7989 & 395.5207 \\
		Cow & Fib. cont. & 3 & 11587.7134 & 399.0234 & 400.4352 & 402.1570 \\
		Cow & Fib. disc. & 4 & 10115.2456 & 395.3156 & 397.8156 & 399.4936 \\
		\hline
	\end{tabular}
\end{table}

Overall, these results show that the Fibonacci-based formulation can reproduce growth trajectories across organisms with very different masses and time scales. 
The Fibonacci-based formulation was favored in two of the four cases, while the WBE model was favored in the other two. 
For guppy, the Fibonacci models were favored, although the corrected criterion selected the continuous form rather than the discretized one. 
For hen, the discretized Fibonacci model was favored by the information criteria. 
For guinea pig and cow, the WBE equation remained the best supported model, but the Fibonacci curves still followed the empirical growth patterns closely. 
Therefore, the Fibonacci-based formulation should not be interpreted as a universal replacement for the WBE model. 
Rather, it provides an alternative ontogenetic parametrization that can be competitive in some biological systems, depending on the dataset and on the information criterion considered.

\noindent\textbf{Additional evaluation using rodent growth data.}
As a complementary analysis, the Fibonacci-based discretized model was also tested using rodent growth data. 
This analysis included C57BL/6J mice, Wistar rats, and Sprague Dawley rats, with males and females considered separately. 
The purpose of this additional test was to examine whether the Fibonacci-based representation remains useful when applied to independent growth series.

In the rodent datasets, the Fibonacci discretized model was compared with classical continuous growth models, including Gompertz, logistic, von Bertalanffy, and Richards equations. 
The best established continuous model was selected using the corrected Akaike information criterion. 
The Fibonacci discretized model was then compared with this reference using RMSE, AIC, \(AIC_c\), and BIC.

The results showed that the performance of the Fibonacci discretized model was not uniform across all rodent series. 
In some datasets, such as Wistar females and Sprague Dawley males, the Fibonacci model produced RMSE values close to or lower than the best established continuous model. 
In other datasets, the established continuous models were favored, especially when model complexity was penalized by \(AIC_c\). 
Thus, the rodent analysis supports a cautious interpretation. 
The Fibonacci-based discretized model can be competitive in some growth series, but it does not systematically outperform classical continuous models.

\begin{table}[htbp]
	\centering
	\caption{
		Summary of the complementary rodent analysis using the corrected Akaike information criterion. 
		The ratio corresponds to \(RMSE_{\mathrm{Fib}}/RMSE_{\mathrm{classical}}\). 
		\(\Delta AIC\), \(\Delta AIC_c\), and \(\Delta BIC\) are computed as Fibonacci minus classical.
	}
	\label{tab:rodent_summary_aicc}
	\small
	\begin{tabular}{l l c c c c c c}
		\hline
		Dataset & \makecell{Best \\ continuous} & \makecell{RMSE \\ class.} & \makecell{RMSE \\ Fib.} & Ratio & \(\Delta AIC\) & \(\Delta AIC_c\) & \(\Delta BIC\) \\
		\hline
		\makecell{C57BL/6J \\ male} & von Bertalanffy & 0.8799 & 0.8968 & 1.0192 & 2.8352 & 3.8549 & 3.9263 \\
		\hline
		\makecell{C57BL/6J \\ female} & von Bertalanffy & 0.8273 & 0.8234 & 0.9953 & 1.7927 & 2.8123 & 2.8837 \\
		\hline
		Wistar male & Logistic & 3.4828 & 4.3230 & 1.2413 & 6.3226 & 10.3226 & 6.6252 \\
		\hline
		Wistar female & Gompertz & 1.9155 & 1.7624 & 0.9201 & 0.3340 & 4.3340 & 0.6366 \\
		\hline
		\makecell{Sprague \\ Dawley male} & Gompertz & 6.7874 & 4.6814 & 0.6897 & -5.4292 & -1.4292 & -5.1266 \\
		\hline
		\makecell{Sprague \\ Dawley female} & Gompertz & 2.0893 & 2.0817 & 0.9964 & 1.9270 & 5.9270 & 2.2296 \\
		\hline
	\end{tabular}
\end{table}

Taken together, the WBE comparison and the rodent analysis suggest that the Fibonacci-based discretized model is best viewed as an alternative representation of growth. 
It is not uniformly superior to established models, but it can reproduce empirical trajectories with reasonable accuracy and, in some cases, with information-criterion support comparable to or better than classical references.

\FloatBarrier
\section{Conclusion}

In this work, we tested a Fibonacci-based ontogenetic model for describing body-mass growth trajectories. 
The main comparison was made with the WBE ontogenetic growth model using digitized data for guinea pig, guppy, hen, and cow. 
These organisms represent very different biological systems, with distinct body masses and developmental times.

The results showed that the Fibonacci discretized model was able to reproduce the general shape of the empirical growth trajectories. 
It did not systematically outperform the WBE model. 
However, the Fibonacci-based formulation was favored in two of the four analyzed cases, Guppy and Hen, while the WBE model remained favored for Guinea pig and Cow. 
Even in the cases where the WBE model was favored, the Fibonacci curves still followed the observed growth patterns closely.

A complementary analysis using rodent growth data also showed that the Fibonacci discretized model can be competitive in some datasets. 
In particular, it produced RMSE values close to or lower than those of the best established continuous models in some growth series. 
However, the inclusion of \(AIC_c\) made the rodent comparison more conservative, especially because some series contained a limited number of data points. In several cases, the corrected information criterion favored the simpler established continuous models. 
Therefore, the rodent analysis should be interpreted as complementary support for the flexibility of the Fibonacci-based representation, not as evidence of systematic superiority.

These findings support a cautious interpretation of the proposed model. 
In the present approach, the discretization is applied to the ontogenetic coordinate, not directly to measured body mass. 
Thus, the step-like structure represents a coarse-grained description of developmental progression.

The continuous Fibonacci-based model can be interpreted as a Gompertz-type reparametrization written in terms of an ontogenetic coordinate. 
Therefore, the main contribution of the present approach is not the continuous growth curve alone, but the use of a Fibonacci-inspired coordinate and its discretized form to describe stage-like progression during growth.

Overall, the results suggest that a Fibonacci-based ontogenetic coordinate can provide a useful alternative representation of body-mass growth within a biological systems framework. 
The model is not proposed as a universal replacement for classical growth equations, but as a complementary way to organize and analyze growth trajectories. 
Future studies using original longitudinal datasets, individual-level growth data, and metabolic measurements may help clarify whether this ontogenetic discretization has deeper biological meaning or should be interpreted mainly as an effective mathematical representation.

\end{document}